\magnification=\magstep2
\centerline{\bf Vortex Waves in a Cloud of Bose Einstein - Condensed,} 
\centerline{\bf Trapped Alkali - Metal Atoms}
\bigskip
\bigskip

\centerline{Carlo F. Barenghi}
\bigskip

\centerline{Mathematics Department}
\centerline{University of Newcastle upon Tyne}
\centerline{NE1 7RU England}
\bigskip

\bigskip

\bigskip
\centerline{Abstract}
\bigskip

\noindent
We consider the vortex  state solution for a rotating cloud of trapped, 
Bose Einstein - condensed alkali atoms and study finite temperature
effects. We find that thermally excited vortex waves can
distort the vortex state significantly, even at the very low temperatures
relevant to the experiments.
\bigskip

\bigskip

\bigskip
\noindent
PACS numbers 0.35 Fi, 5.30 Jp, 32.80 Pj
\vfill
\eject

Recent experiments [1,2,3] have demonstrated the existence of Bose - Einstein 
condensation  [4,5,6] in clouds of trapped alkali atoms and generated
great theoretical interest [7,8]. To understand the superfluid
aspects of these new systems it is natural to study what happens when
the clouds are rotated [9]. 
The existence of quantised vortex lines in rotating clouds 
has not been experimentally established yet. Nevertheless
a vortex state solution of the governing Gross - Pitaevskij equation 
has been found already by Dalfovo and Stringari [10]. They solved the
equation for systems of $N \approx 10^4$ atoms at temperature $T=0$ 
flowing with quantised circulation around a vortex line set on the $z$ 
axis of rotation of the cloud. They also argued that the new solution is 
relevant to the stability [11] of trapped clouds of atoms which have negative 
scattering length, such as $^7$Li [3]. Dalfovo and Stringari's work
has extended to new systems the study of quantised vorticity, until now 
restricted to superfluid helium and neutron stars [12], a
development which is important as new non - destructive detection methods
are invented [13].
\bigskip

The aim of this brief report is to estimate some
finite temperature corrections. We find that the vortex state calculated 
at $T=0$  by Dalfovo and Stringari can be rather distorted at
small but finite $T$ by sinusoidal displacements of the vortex core 
away from the axis of rotation. These displacements, 
called {\it vortex waves} [14,15], are excited thermally. The effect 
is at first surprising, given the very low temperatures at which clouds 
of trapped Bose Einstein -  condensed atoms are produced, about $10^{-7}K$ 
only. 
\bigskip

Let us consider a cloud of size $L=2 R_c$ of trapped, Bose Einstein - 
condensed atoms at temperature $T$. The cloud rotates about the $z$ 
axis at angular velocity $\Omega \ge \Omega_{crit}$, where the critical 
velocity $\Omega_{crit}$ for the appearance of the first vortex line
is typically of the order of few Hertz [6,10]. 
The vortex line is located on the $z$ axis in its unperturbed state
at $T=0$. Let us assume that at temperature $T>0$ the vortex core is 
displaced away from the axis of rotation by the amount

$$\eta(z,t)=A \cos{(\omega t)} \cos{(k z)},\eqno{(1)}$$ 

\noindent
where $\lambda$, $k=2 \pi/\lambda$, $\omega$ and $A$ are respectively
the wavelength, the wavenumber, the angular frequency and the amplitude
of the wave. In the first approximation we assume that the cloud is 
large enough that we can neglect size effects [16] on the dispersion 
relation $\omega=\omega(k)$ of the wave. We have then
$\omega=\Gamma k^2 {\cal L}/4 \pi$ [12],
where $\Gamma=h/m$ is the quantum of circulation, $h$ is Plank's 
constant, $m$ is the mass, ${\cal L } \approx \ln{ (2/k a_{core})}$
and $a_{core}$ is the vortex core parameter.
To determine the amplitude of a vortex wave of wavenumber $k$ we note 
that it is a linear
perturbation of the vortex state solution of the Gross - Pitaevskij 
equation [17], in the same way as phonons are perturbations of the uniform
solution. Assuming that vortex waves are quantized like harmonic 
oscillators, the energy ${\cal E}$ of the wave of wavenumber $k$ at 
temperature $T$ is equal to the energy $\hbar \omega$ times the mean number 
$1/[\exp{(\beta \hbar \omega)}-1]$ of excitations in that state $k$.
There are two contributions to
${\cal E}$: the kinetic energy ${\cal E}_1$ of the fluid rotating around 
the core and the potential energy ${\cal E}_2$ due to the increase 
$\delta \ell$ of the vortex' length from the unperturbed to the
pertuebed state. We have

$${ { \hbar \omega }\over{e^{\beta \hbar \omega}-1}}
={\cal E}_1 + {\cal E}_2,\eqno{(1)}$$

\noindent
where $\beta=1/K_B T$ and $K_B$ is Boltzmann's constant. The kinetic 
energy is obtained from averaging the energy density
$\rho_{eff} \pi a_{core}^2 \dot{\eta}^2/2$ 
over the period $2 \pi/\omega$ of oscillation and the dimension of the 
system, where $\rho_{eff} = \rho/2$ is the hydrodynamic mass and 
$a_{core}$ is the size of the vortex core:

$${\cal E}_1={{\pi}\over{8}} a_{core}^2 \rho_{eff} A^2 \omega^2 L
,\eqno{(2)}$$ 

\noindent
The potential energy is ${\cal E}_2=\epsilon \delta \ell$
where $\epsilon$ is the energy of the vortex motion per unit length in
the $z$ direction, $\ell=L$ is the length of the unperturbed vortex and 
$\ell'=\int_0^L dz \sqrt{1+A^2 k^2 \cos^2{(kz)}}$ 
is the length of the vortex in the presence of waves. To obtain $\epsilon$ 
we integrate $\rho v^2/2$ around the vortex line, 
where $v=\Gamma/2 \pi r$ is found from the quantization of circulation. 
We obtain

$${\cal E}_2=\delta \ell { {\rho \Gamma^2 } \over {4 \pi} }
\ln{(R_c/a_{core})},\eqno{(2)}$$

\noindent
The largest amplitudes arise from the  smallest wavenumbers, so we 
assume $\lambda \approx L=2 R_c$. In the limit $A k << 1$ we can 
Taylor expand the integrand for $\ell'$. We also note that typically 
$\beta \hbar \omega << 1$, and, neglecting a term proportional to 
$a_{core}^2/R_c^2 <<1$, we conclude that

$$A \approx \sqrt{{32 K_B T R_c}\over
{\pi \rho \Gamma^2 ln{(R_c/a_{core})}}},
\eqno{(4)}$$

\noindent
We apply equation (4) to the $^{87}$Rb experiments of Ref. 1 and the 
discussion  of Dalfovo and Stringari [10]. The quantum of circulation
is $\Gamma=4.4 \times 10^{-5}~cm^2/sec$. The average density of atoms 
in typical clouds is estimated to be in the range from 
$n \approx 10^{12}$ to $10^{13}~cm^{-3}$ which corresponds to
$\rho \approx 10^{-10}$ to $10^{-9}~g/cm^3$. 
The vortex core parameter enters equation (4) only 
via a slow logaritmic term. From the calculation of Dalfovo and Stringari 
for a small cloud of $N=5,000$ atoms (their figure 4b) we infer 
that $a_{core}$, which we define for convenience as the distance from the 
axis over which the computed condensate's wave function recovers half of 
its peak value, is $a_{core} \approx 0.2 a_{\perp}$ 
$\approx 0.24 \times 10^{-4}~cm$ where 
$a_{\perp} \approx 1.2 \times 10^{-4}~cm$ is the oscillator's 
characteristic length. 
At $T=10^{-7}~K$ and $R_c \approx 5 \times 10^{-4}~cm$ [6,10]
we find that $A$ ranges from $0.33 \times 10^{-3}$
to $0.11 \times 10^{-3}~cm$. This amplitude is not a negligible 
distortion of the vortex state: it represents a displacement of the 
vortex line of a distance from $4$ to $14$ times the size of the core. 
At larger values of $R_c$ we expect size effects to become less important
for which equation (4) becomes a better estimate of $A$.
For example at $R_c \approx 10^{-3}~cm$
the displacements  are $6$ to $20$ times larger than $a_{core}$.
It is instructive to compare this result with the case of a vortex line 
in superfluid $^4$He. In $^4$He the vortex waves are negligible. Using
$\rho = 0.145~g/cm^3$ and $a_{core} \approx 10^{-8}~cm$ [18] and the 
same $R_c=5 \times 10^{-3}~cm$ and $T=10^{-7}~K$ used before, we find
$A \approx 0.2 \times 10^{-9}~cm$, a displacement which is only two 
hundredths of helium's vortex core size.
\bigskip

Finally we compare the thermally induced displacements
of the core with the amplitude of the zero - point motion 
of the vortex line, $A_0 \approx 0.5 \sqrt{2 \hbar /\rho \Gamma a_{core}}$
[19]. Using the two values of $\rho$ mentioned above we find that 
$A_0/a_{core}$ ranges from $1$ to $3$, which is less than the  
ratio $A/a_{core}$ for the thermal displacements. For vortex lines in
$^4$He one has $A_0/a_{core} = 2$.
\bigskip

We conclude that long wavelength thermal waves can affect the vortex 
state even at the very low temperatures relevant to the experiments.
\vfill
\eject

\centerline{\bf References}
\bigskip

\noindent
[1] M.H. Anderson, J.R. Ensher, M.R. Matthews, C.E. Wieman \& 
E.A. Cornell, Sciemce {\bf 269}, 198 (1995).
\bigskip

\noindent
[2] K.B. Davis, M.O. Mewes, N.J. van Druten, D.S. Durfee,
D.M. Kurn \& W. Ketterie, Phys. Rev. Letters {\bf 75}, 3969 (1995).
\bigskip

\noindent 
[3] C.C. Bradley, C.A. Sackett, J.J. Tollett \& R.G. Hulet,
Phys. Rev. Lett. {\bf 75}, 1687 (1995).
\bigskip

\noindent
[4] A. Griffin, D.W. Snoke \& S. Stringari ed's. {\it Bose Einstein
Condensation} Cambridge University Press (1996).
\bigskip

\noindent
[5] A.L. Fetter, Phys. Rev. A {\bf 53}, 4245 (1996).
\bigskip

\noindent 
[6] G. Baym \& C.J. Pethick, Phys. Rev. Letters {\bf 76}, 6 (1996).
\bigskip

\noindent
[7] M. Edwards \& K. Burnett, Phys. Rev. A {\bf 51}, 1382 (1995).
\bigskip

\noindent
[8] P.A. Ruprecht, M.J. Holland, K, Burnett \& M. Edwards,
Phys. Rev. A {\bf 51}, 4704 (1995).
\bigskip

\noindent
[9] S. Stringari, Phys. Rev. Letters {\bf 76}, 1405 (1996).
\bigskip

\noindent 
[10] F. Dalfovo \& S Stringari, Phys. Rev. A {\bf 53}, 2477 (1996).
\bigskip

\noindent 
[11] L.P. Pitaevskij, preprint cond-mat/9605119 (1996).
\bigskip

\noindent 
[12] R.J. Donnelly, {\it Quantized Vortices in Helium II},
Canbridge University Press (1992).
\bigskip

\noindent 
[13] M.R. Andrews, M.O. Mewes, N.J. Vandruten, D.S. Durfee,
D.M. Kurn \& W. Ketterle, Science {\bf 273}, 84 (1996).
\bigskip

\noindent 
[14] W.I. Glaberson \ R.J. Donnelly, in {\it Progress in Low 
 Temperature Physics} vol. IX, edited by D.F.Brewer, Elsevier Science
 Publishers, page 1 (1986)
\bigskip

\noindent 
[15] C.F. Barenghi, R.J. Donnelly \ W.F. Vinen,
Phys. Fluids {\bf 28}, 498 (1985).
\bigskip

\noindent
[16] In the case of phonons, size effects for small clouds
have been calculated
by S. Stringari, preprint cond-mat/9603126 (1996) and
M. Edwards, P.A. Ruprecht, K. Burnett, R.J. Dodd \&
C.W. Clark, preprint cond-mat / 9605170 (1996).
\noindent

\noindent
[17]  L.P. Pitaevskij, Sov. Phys. JETP {\bf 13}, 451 (1961)
\bigskip

\noindent 
[18] C.F. Barenghi, R.J. Donnelly \& W.F. Vinen,
J. Low Temp. Phys. {\bf 52}, 189 (1982).
\bigskip

\noindent 
[19] A.L. Fetter, Phys. Rev. {\bf 162}, 143 (1967).
\bigskip

\bye